\documentstyle[12pt,psfig]{article}

\begin{document}
\title{ $ee^+\rightarrow\pi^0\gamma$ and form factor of $\pi^0\gamma^*\gamma^*$}
\author{Bing An Li\\
Department of Physics and Astronomy, University of Kentucky\\
Lexington, KY 40506, USA}

\maketitle
\begin{abstract}
The form factor $\pi^0\gamma^*\gamma^*$ is obtained to the
next leading order of derivative expansion of the chiral anomaly and 
the VMD. As a test the form factor $\pi^0\gamma^*\gamma$ has been used
to calculate the cross section of $ee^+\rightarrow\pi^0\gamma$. Theory agrees with data 
well. 
\end{abstract}
\newpage

A measurement of the muon g-2 with high accuracy has been reported[1] as
\[a_\mu=11659202(14)(6)\times10^{-10}.\]
The hadronic contributions to the muon g-2 consist of vacuum polarization, 
high order
corrections, and light-by-light scattering. The contribution of light-by-light 
scattering to the muon g-2($a^{lbl}$) has been studied[2,3]. The sign problem 
no longer exists.
The pseudoscalar poles play a dominant roles, especially, the pion pole.
Various transition form factor of $\pi^0\gamma^*\gamma^*$ have been
used in the calculation of $a^{lbl}$. The Wess-Zumino-Witten term, 
Vector Meson Dominance, and ENJL model are all used to obtain the transition
form factor of $\pi^0\gamma^*\gamma^*$. In Ref.[3] four different form factors
of $\pi^0\gamma^*\gamma^*$ are used to calculate the contribution to 
$a^{lbl}$.

The measurements of the form factor $\pi^0\gamma\gamma^*$ with one photon 
on mass 
shell  
have lasted for a long time[4,5].
In the timelike region the slope of the form factor of
$\pi^0\rightarrow\gamma e^+ e^-$
\[F(q^2)=1+a\frac{m^2_{e^+ e^-}}{m^2_{\pi^0}}\]
has been measured[5]. 
The measured value of the slope falls in a wide range of -0.24 to 0.12.
CELLO and CLEO[4] have measured the form factors of $P\gamma\gamma^*$ 
in the range of 
large $q^2$.
The PrimEx Coll. of JLab
is going to do direct precision measurements of the form factor of
$\pi^0\gamma\gamma^*$ at small values of $q^2$,
$0.001GeV^2 \leq q^2 \leq 0.5 GeV^2$[6].
On the other hand,
the form factor of $\pi^0\gamma\gamma^*$ has been studied by various
theoretical approaches[7].

The form factor of $\pi^0\gamma^*\gamma^*$ is related to the chiral anomaly.
When two photons are on mass shell in the chiral limit the vertex $\pi^0\gamma\gamma$
is the Adler-Bell-Jackiw anomaly(ABJ)[8] and the amplitude is
\begin{equation}
A_\pi=\frac{2\alpha}{\pi f_\pi}.
\end{equation}
In the chiral limit the form factor $\pi^0\gamma^*\gamma^*$ is via the 
Vector Meson Dominance(VMD) derived from the anomalous vertex $\pi\omega\rho$. 
This vertex is obtained from  
the Bardeen form of the Wess-Zumino-Witten(WZW) anomalous Lagrangian of pseudoscalar,
vector, and
axial-vector fields, which has been presented by Kaymakcalan, Rajeev, and Schechter
(KRS)[9]
\begin{eqnarray}
\lefteqn{{\cal L}=
\frac{N_{c}}{(4\pi)^{2}}{2\over 3}\varepsilon^{\mu\nu\alpha\beta}
\omega_{\mu}Tr\partial_{\nu}UU^{\dag}\partial_{\alpha}UU^{\dag}
\partial_{\beta}UU^{\dag}}\nonumber \\
 & &+\frac{2N_{c}}{(4\pi)^{2}}\varepsilon^{\mu\nu\alpha\beta}
\partial_{\mu}\omega_{\nu}Tr\{i[\partial_{\beta}UU^{\dag}
(\rho_{\alpha}+a_{\alpha})-\partial_{\beta}U^{\dag}U(\rho_{\alpha}
-a_{\alpha})]\nonumber \\
& &-(\rho_{\alpha}+a_{\alpha})U(\rho_{\beta}-a_{\beta})
U^{\dag}-2\rho_{\alpha}a_{\beta}\},\\
& &{\cal L}_{\pi\omega\rho}=-\frac{3}{\pi^2 g^2 f_\pi}\pi_i
\varepsilon^{\mu\nu\lambda\beta}
\partial_\mu\rho^i_\nu\partial_\lambda\omega_\beta,
\end{eqnarray}
where g is the universal coupling constant which is introduced by normalizing 
the vector fields
\[\rho^i_\mu\rightarrow {1\over g}\rho^i_{\mu},\;\;\;
\omega_\mu\rightarrow {1\over g}\omega_{\mu}.\]
The universal coupling constant g appears in the VMD[10]
\begin{eqnarray}
\lefteqn{{1\over2}eg\{-{1\over2}F^{\mu\nu}(\partial_\mu \rho_\nu
-\partial_\nu \rho_\mu)+A^\mu j_\mu\}},\nonumber \\
&&{1\over6}eg\{-{1\over2}F^{\mu\nu}(\partial_\mu \omega_\nu
-\partial_\nu \omega_\mu)+A^\mu j^0_\mu\}.
\end{eqnarray}
The ABJ anomaly is derived from Eqs.(3,4).

The processes contributing to the form factor $\pi^0\gamma^*\gamma^*$ are shown in
Fig.1(a-d) and the matrix element is expressed as
\begin{equation}
<\gamma_1\gamma_2|S|\pi^0>=-i(2\pi)^4\delta^4(p-q_1-q_2)\frac{1}
{\sqrt{8m_\pi\omega_1\omega_2}}
\varepsilon^{\mu\nu\lambda\beta}\epsilon_\mu(1)\epsilon_\nu(2)q_{1\lambda}q_{2\beta}
\frac{2\alpha}{\pi f_\pi}F(q^2_1,q^2_2).
\end{equation}
Using Eqs.(3,4), the vertices in Fig.1 are obtained and the form factor is determined 
\begin{eqnarray}
F(q^2_1,q^2_2)& = &{1\over2}
\{\frac{m^2_\rho m^2_\omega}{(q^2_1-m^2_\rho)(q^2_2-m^2_\omega)}
+\frac{m^2_\rho m^2_\omega}{(q^2_1-m^2_\omega)(q^2_2-m^2_\rho)}\},
\end{eqnarray}
where $q_1, q_2$, and p are momentum of two photons and pion 
respectively. Eq.(6) shows that the form factor is dominated by the poles of vector 
mesons. This form factor is determined by the Bardeen form of the WZW anomaly 
presented by KRS and the VMD.

It is necessary to point out that in effective chiral theory of mesons derivative
expansion is taken as the low energy approximation. 
The chiral anomalous Lagrangian(2) is at the $4^{th}$ order(the lowest order) 
of covariant derivative expansion. According to Eq.(6), the slope of the form factor 
is determined by the masses of the vector mesons only. In principle, the chiral anomaly
should include terms at higher orders in covariant derivatives. Of course, in the
chiral limit the terms at higher orders don't contribute to $\pi^0\gamma\gamma$(two
photons are on mass shell). However, the terms at higher order contribute to 
${\cal L}_{\pi\omega\rho}$, therefore, they contribute to $F(q^2_1,q^2_2)$. 
We use the form factor of charged pion 
to illustrate the correction of the $\rho$ pole-form factor. It is known that the  
$\rho$ pole-form factor of charged pion has problems:
the radius is less than data by about $10\%$, in the space-like region the form factor 
decreases too fast and in the time-like region it decreases too slow[11]. In Ref.[12]
by calculating the next leading order in derivative expansion besides the $\rho$ pole
an intrinsic form factor 
\begin{equation}
f_{\rho\pi\pi}(q^2)=1+\frac{q^2}{2\pi^2 f^2_\pi}\{(1-{2c\over g})^2
-4\pi^2 c^2\}
\end{equation}
has been found, where 
\[c=\frac{f^2_\pi}{2gm^2_\rho}.\]  
The intrinsic form factor $f_{\rho\pi\pi}$ remedies these problems. The form
factor $F_\pi(q^2)$ agrees with data in both space-like and time-like 
regions($q^2
\sim 1.4GeV^2$).  
On the other hand, the radius of charged pion
obtained from $\rho$ pole is
\[<r^2>_\pi=0.395 fm^2\]
and with the contribution of the additional form factor(7) we obtain
\[<r^2>_\pi={6\over m^2_\rho}+
\frac{3}{\pi^2 f^2_\pi}\{(1-{2c\over g})^2-4\pi^2 c^2\}=
0.452 fm^2.\]
The data is $(0.44\pm0.03)fm^2$[13]. $13\%$ of the radius comes from the intrinsic
form factor $f_{\rho\pi\pi}$. The contribution of the next 
leading order to the slope is at the same order as the one from the $\rho$-pole.

The effects of next leading order of derivative expansion on the form factor 
$\pi^0\gamma^*\gamma^*$ needs to be studied.

We have proposed an effective chiral theory of large $N_C$ QCD of
pseudoscalar, vector, and axial-vector mesons[14].
In the limit $m_q\rightarrow 0$, this theory is chiral symmetric and
has dynamical chiral
symmetry breaking. Based on the current algebra the Lagrangian is constructed as
\begin{eqnarray}
{\cal L}=\bar{\psi}(x)(i\gamma\cdot\partial+\gamma\cdot v
+\gamma\cdot a\gamma_{5}+eQ\gamma\cdot A
-mu(x))\psi(x)-\bar{\psi(x)}M\psi(x)\nonumber \\
+{1\over 2}m^{2}_{0}(\rho^{\mu}_{i}\rho_{\mu i}+
\omega^{\mu}\omega_{\mu}
+a^{\mu}_{i}a_{\mu i}
+f^{\mu}f_{\mu})
\end{eqnarray}
where M is the quark mass matrix
\[\left(\begin{array}{c}
         m_{u}\hspace{0.5cm}0\\
         0\hspace{0.5cm}m_{d}
        \end{array}  \right ),\]
\(v_{\mu}=\tau_{i}\rho^{i}_{\mu}
+\omega_{\mu}\)
,
\(a_{\mu}=\tau_{i}a^{i}_{\mu}
+f_{\mu}\),
\(u=exp\{i\gamma_{5}(\tau_{i}\pi_{i}+
\eta)\}\), and m is the constituent quark mass which is related to dynamical 
chiral symmetry breaking.
The kinetic terms of mesons are generated by quark loops.
By integrating out the quark fields, the Lagrangian of mesons is
obtained. As shown in Ref.[14] the effective Lagrangian of mesons has real and
imaginary two parts. All the vertices of chiral anomaly are from the imaginary part and
the normal vertices are from the real part. Like all other effective meson theories,
the derivative expansion is taken as a low energy approximation. At the 
leading order($4^{th}$)
of derivative expansion the imaginary part of the Lagrangian 
obtained from Eq.(8) is exact the same
as the Bardeen form of the WZW anomaly presented By KRS(2)(see Eqs.(98-99) of Ref.[14]).
As a matter of fact, the fields in the Bardeen form of the WZW anomaly presented by KRS
needs to be normalized to physical meson fields. The normalizations are carried 
out by the kinetic terms of the fields of the real part of the Lagrangian, for example,
the normalizations of $\rho$ and $\omega$ fields mentioned above and
\[\pi\rightarrow{2\over f_\pi}\pi.\]
The universal coupling constant g and $f_\pi$ are defined as
\begin{eqnarray}
\lefteqn{g^2=\frac{8N_C}{3(2\pi)^4}\int\frac{d^4 k}{(k^2+m^2)^2},}\nonumber \\
&&f^2_\pi(1-{2c\over g})^{-1}=
\frac{16m^2N_C}{(2\pi)^4}\int\frac{d^4 k}{(k^2+m^2)^2}=6m^2 g^2.
\end{eqnarray}
In the chiral limit g and $f_\pi$ are two inputs. 
g is determined
to be 0.39 by fitting
the decay rate of $\rho\rightarrow ee^+$ and
we take \(f_\pi=0.186GeV\).
$N_C$ expansion is revealed from this theory.
The tree diagrams are at leading order of $N_C$
expansion and loop diagrams of mesons are at higher orders. At low energies 
this theory goes back to the 
Chiral Perturbation Theory(ChPT) and the 10 coefficients of ChPT are 
determined[15].
The VMD is a natural result of this theory. In this theory both the anomalous vertices
(2)and the VMD(10) are derived from the same Lagrangian(8). In the chiral limit there are 
two parameters g and $f_\pi$ which have been fixed.   
Many physics processes have been calculated and theory agrees with data well
[12,14,16].
The theory is phenomenological successful. The form factor of charged pion 
mentioned above is the result of this theory. 

In this paper we use the effective chiral theory(8) to study 
the contributions of the terms at the $6^{th}$ order in derivative expansion to 
the form factor $\pi^0\gamma^*\gamma^*$ at energies
up to about 1 GeV.
The vertices of the processes shown in Fig.1 to next leading order in derivative 
expansion 
can be obtained from the imaginary part of the Lagrangian(1) by calculating 
corresponding quark loops. 
The related vertices to the sixth order in derivatives are found 
\begin{eqnarray}
\lefteqn{{\cal L}_{\pi^0\gamma\gamma}=-\frac{e^2}{4\pi^2 f_\pi}
\{1+{g^2\over2f^2_\pi}(1-{2c\over g})^2
(q^2_1+q^2_2+p^2)\}\pi^0\varepsilon^{\mu\nu\lambda\beta}\partial_\mu A_\nu
\partial_\lambda A_\beta,}\\
&&{\cal L}_{\pi^0\rho\gamma}=-\frac{e}{2g\pi^2 f_\pi}
\{1+\frac{g^2}{2f^2_\pi}(1-{2c\over g})^2 (q^2_1+q^2_2+p^2)\}
\pi^0\varepsilon^{\mu\nu\lambda\beta}\partial_\mu A_\nu
\partial_\lambda \rho_\beta,\\
&&{\cal L}_{\pi^0\omega\gamma}=-\frac{3e}{2g\pi^2 f_\pi}
\{1+\frac{g^2}{2f^2_\pi}(1-{2c\over g})^2(q^2_1+q^2_2+p^2)\}
\pi^0\varepsilon^{\mu\nu\lambda\beta}\partial_\mu A_\nu
\partial_\lambda \omega_\beta,\\
&&{\cal L}_{\rho\gamma}=-{e\over4}g(\partial_\mu
A_\nu-\partial_\nu A_\mu)
\{1-{1\over10\pi^2 g^2}{\partial^2\over m^2}\}(\partial^\mu \rho^{0\nu}
-\partial_\nu
\rho^{0\mu}),\\
&&{\cal L}_{\omega\gamma}=-{e\over12}g(\partial_\mu A_\nu-\partial_\nu A_\mu)
\{1-{1\over10\pi^2 g^2}{\partial^2\over m^2}\}(\partial^\mu \omega^{\nu}
-\partial^\nu
\omega^{\mu}),\\
&&{\cal L}_{\pi^0\omega\rho}=-{3\over \pi^2 g^2 f_\pi}\pi^0
\varepsilon^{\mu\nu\lambda\beta}
\partial_\mu\rho^0_\nu\partial_\lambda\omega_\beta\{1+\frac{g^2}{2f^2_\pi}
(1-{2c\over g})^2
(q^2_1+q^2_2+p^2)\},
\end{eqnarray}
where $q_1^2$, $q_2^2$, and $p^2$ are momentum of $\rho(\gamma)$, 
$\omega(\gamma)$, and $\pi^0$
respectively.
In the chiral limit, if the two photons are on mass shell Eq.(10) goes back to 
the ABJ anomaly
\begin{equation}
{\cal L}_{\pi^0 \rightarrow\gamma\gamma}=-\frac{\alpha}{\pi f_\pi}
\varepsilon^{\mu\nu\lambda\beta}\pi^0 \partial_\mu A_\nu \partial_\lambda
A_\beta.
\end{equation}
Up to the $6^{th}$ order in derivatives,
the form factor of $\pi^0\gamma^*\gamma^*$(5) is determined to be
\begin{equation}
F(q^2_1,q^2_2)={1\over2}
\{\frac{m^2_\rho m^2_\omega}{(q^2_1-m^2_\rho)(q^2_2-m^2_\omega)}
+\frac{m^2_\rho m^2_\omega}{(q^2_1-m^2_\omega)(q^2_2-m^2_\rho)}
+\frac{g^2}{2f^2_\pi}(1-{2c\over g})^2(q^2_1+q^2_2+p^2)\},
\end{equation}
in space-like region,
and $q^2_1$, $q^2_2$, and $p^2$ are momentum of two photons and pion 
respectively. Comparing with Eq.(6), in Eq.(17) there is a new term which is from the 
the terms at the $6^{th}$ order in derivatives.  

Put one photon on mass shell($\pi^0$ too), in the chiral limit 
the form factor of $\pi^0\gamma\gamma^*$ is 
derived from Eq.(17)
\begin{equation}
F_{\pi^0 \gamma\gamma^*}(q^2) = {1\over2}
\{\frac{m^2_\rho}{m^2_\rho-q^2}+\frac{m^2_\omega}{m^2_\omega-q^2}
+\frac{g^2}{2f^2_\pi}(1-{2c\over g})^2 q^2 \}.
\end{equation}
At very low energies the slope of the form factor is obtained
\begin{eqnarray}
F_{\pi\gamma\gamma^*}(q^2) & = & 1+a{q^2\over m^2_\pi},\\
&&a=
{m^2_\pi\over2}({1\over m^2_\rho}+{1\over m^2_\omega})
+{m^2_\pi\over2f^2_\pi}g^2(1-{2c\over g})^2.
\end{eqnarray}
\begin{equation}
a=0.0303+0.0157=0.046.
\end{equation}
The first number of Eq.(21) is from the poles of vector mesons and the second 
number is the contribution of 
the term at the $6^{th}$ order in derivatives, which is 
$34\%$ of the slope.
The value of the slope is the prediction of this theory and is a test of the 
form factor(18).
 
The cross section of $ee^+\rightarrow\pi^0\gamma$ has been measured[17].
$\sigma(ee^+\rightarrow\pi^0\gamma)$ is determined by the time-like form factor 
$\pi^0\gamma^*\gamma$ 
\begin{equation}
\sigma(ee^+\rightarrow\pi^0\gamma)=\frac{\alpha^3}{6\pi^2 f^2_\pi}
(1-{m^2_\pi\over q^2})^3|F(q^2)_{\pi^0\gamma\gamma^*}|^2.
\end{equation}
The cross section(22) is a test of the form factor $\pi^0\gamma^*\gamma^*$
 obtained in this paper.
The time-like form factor in Eq.(22) can be obtained from Eq.(18). 
\begin{equation}
F_{\pi^0\gamma\gamma^*}(q^2)={1\over2}
\{\frac{-m^2_\rho+i\sqrt{q^2}\Gamma_\rho(q^2)}{q^2-m^2_\rho+i\sqrt{q^2}
\Gamma_\rho(q^2)}
+\frac{-m^2_\omega+i\sqrt{q^2}\Gamma_\omega(q^2)}{q^2-m^2_\omega+i\sqrt{q^2}
\Gamma_\omega}+\frac{g^2}{2f^2_\pi}(1-{2c\over g})^2 q^2\},
\end{equation}
where $\Gamma_\omega=8.44MeV$ is taken. The decay width of the wide resonance of $\rho$
is expressed as[12]
\begin{equation}
\Gamma_\rho(q^2)=\frac{\sqrt{q^2}}{12\pi g^2}f^2_{\rho\pi\pi}(q^2)
(1-{4m^2_\pi\over q^2})^{{3\over2}}.
\end{equation}
At $q^2=m^2_\rho$ $\Gamma_\rho=150MeV$ which agrees with data very well. If 
$q^2 > 4m^2_K$ the KK channels are open. However, in the range $q^2< 1GeV^2$ the 
contribution of KK channels to $\Gamma_\rho$ can be ignored.

In time-like region because of the effects of narrow resonances the $\rho-\omega$ 
and $\omega-\phi$ mixings have to be taken into account.  
In the effective 
chiral theory[14] in the leading order of $N_C$ expansion the masses of $\rho$ and 
$\omega$ are degenerated. Therefore, in the leading order of $N_C$ expansion the
$\rho-\omega$ mixing is kinetic[18]
\begin{equation}
{\cal L}_{\rho\omega}=\{-{1\over4\pi^2 g^2}{1\over m}(m_d-m_u)+{1\over 24}e^2 g^2\}
(\partial_\mu \rho_\nu-\partial_\nu \rho_mu)
(\partial_\mu \omega_\nu-\partial_\nu \omega_mu).
\end{equation}
In Ref.[18] Eq.(25) has been used to study $ee^+\rightarrow\pi^+\pi^-$. Theory agrees 
with data well. $m_d-m_u$ is determined to be $4.24\pm0.32$ MeV which is in good 
agreement with the one presented by Leutwyler[19].
In the effective chiral theory[14] nondiagonal element of mass matrix of $\omega$ and 
$\phi$ cannot be generated. However, kinetic mixing exists. We take the $\omega-\phi$
mixing as
\begin{equation}
{\cal L}_{\omega\phi}={1\over2}st
(\partial_\mu \phi_\nu-\partial_\nu \phi_\mu)
(\partial_\mu \omega_\nu-\partial_\nu \omega_\mu),
\end{equation}
where st is a parameter. Taking ${\cal L}_{\rho\omega}$ and ${\cal L}_{\omega\phi}$
into account, the form factor $\pi^0\gamma^*\gamma$ is expressed as
\begin{eqnarray}
\lefteqn{
F_{\pi^0\gamma\gamma^*}(q^2)={1\over2}
\{\frac{-m^2_\rho+i\sqrt{q^2}\Gamma_\rho(q^2)}{q^2-m^2_\rho+i\sqrt{q^2}
\Gamma_\rho(q^2)}
+\frac{-m^2_\omega+i\sqrt{q^2}\Gamma_\omega(q^2)}{q^2-m^2_\omega+i\sqrt{q^2}
\Gamma_\omega}+\frac{g^2}{2f^2_\pi}(1-{2c\over g})^2 q^2}\nonumber \\
&&+\sqrt{2}st\frac{q^2}{q^2-m^2_\phi+i\sqrt{q^2}\Gamma_\phi}
\frac{q^2}{q^2-m^2_\omega+i\sqrt{q^2}\Gamma_\omega}\}\nonumber \\
&&+{10\over3}\{
{1\over4\pi^2 g^2}{1\over m}(m_d-m_u)-{1\over 24}e^2 g^2\}\frac{q^4}{
(q^2-m^2_\rho+i\sqrt{q^2}\Gamma_\rho(q^2))
(q^2-m^2_\omega+i\sqrt{q^2}
\Gamma_\omega)}.
\end{eqnarray}
$st=0.0167$ is chosen to fit the cross section of $ee^+\rightarrow\pi^0\gamma$ at 
$q^2=m^2_\phi$[20]. The results are shown in Fig.2. 
Theory agrees with
data well.

To conclude, the form factor of $\pi^0\gamma^*\gamma^*$ has been found up to next
leading order in derivative expansion. The slope of the form factor is predicted. 
The cross section of $ee^+\rightarrow\pi^0\gamma$ has been computed. Theory agrees 
well with data up to $q^2\sim 1GeV^2$.

This study is supported
by a DOE grant.

\pagebreak
\begin{flushleft}
{\bf Figure Captions}
\end{flushleft}
{\bf FIG. 1.} Processes contributing to $\pi^0\gamma^*\gamma^*$

{\bf FIG. 2.} Cross section of $ee^+\rightarrow\pi^0\gamma$

\begin{figure}
\begin{center}
\psfig{figure=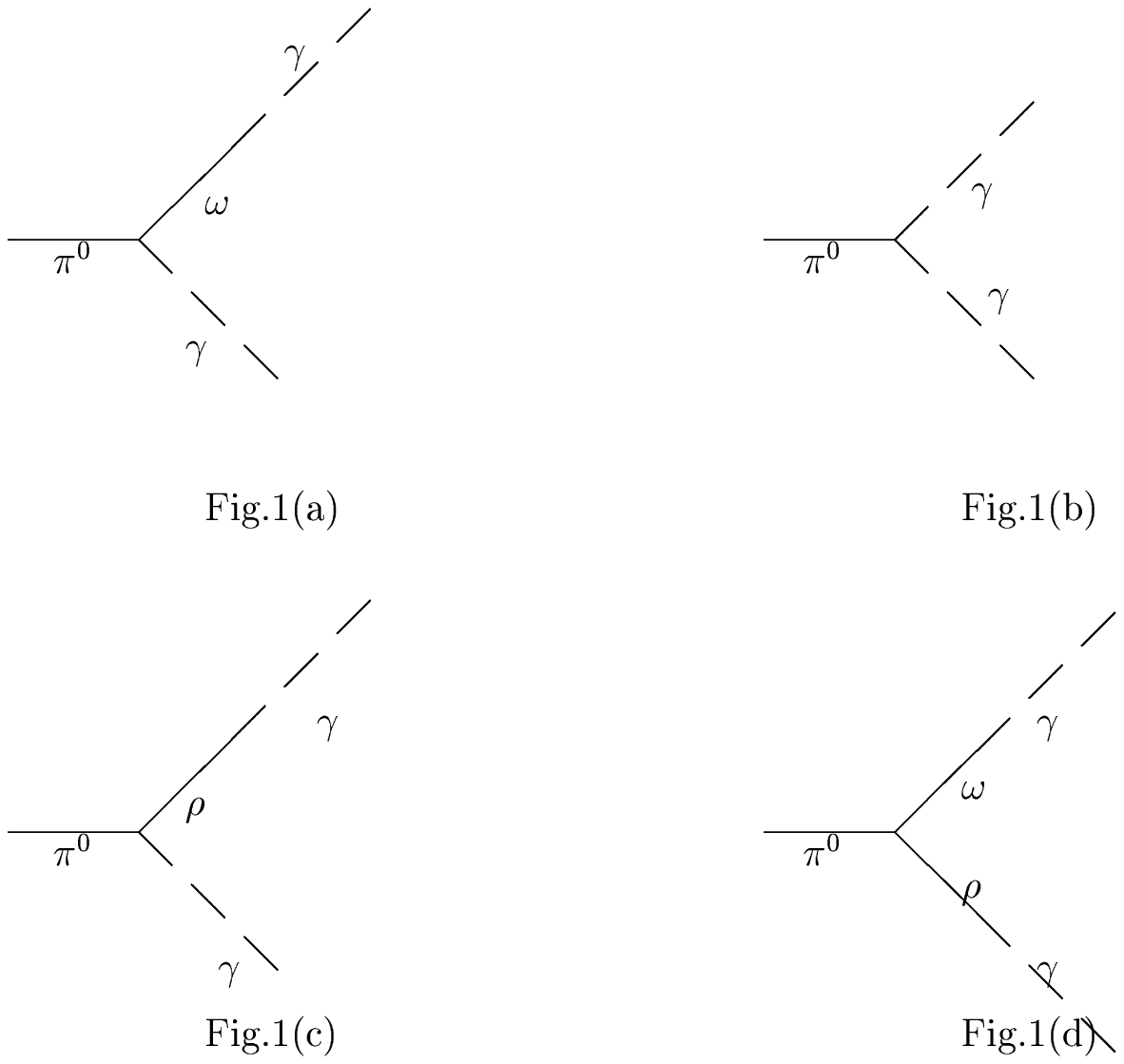}
FIG. 1.
\end{center}
\end{figure}
\begin{figure}
\begin{center}
\psfig{figure=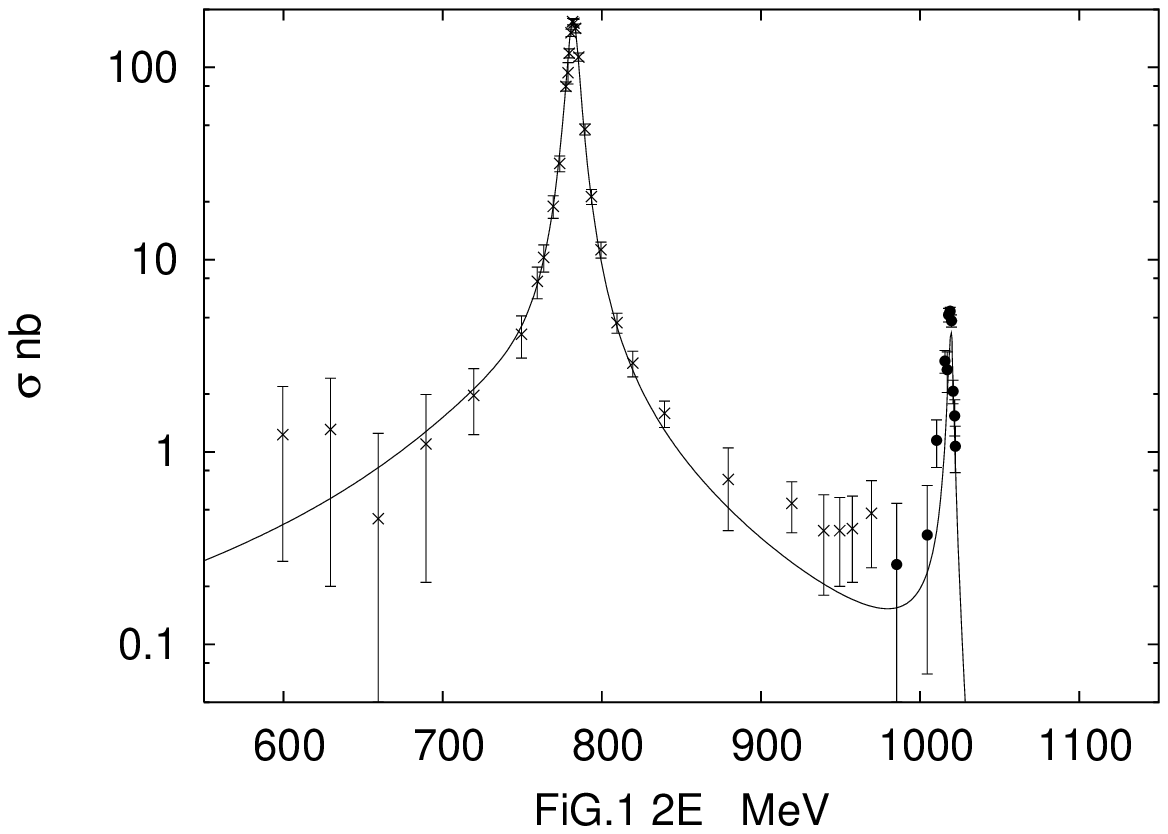}
FIG. 2.
\end{center}
\end{figure}


\begin{thebibliography}{30}
\bibitem{} I have used Eq.(26) and the value of st to study $ee^+\rightarrow3\pi$.
Correct result for $\phi\rightarrow3\pi$ is obtained. The study will be presented 
somewhere else.
\end{thebibliography}
\end{document}